# Electrical generation and absorption of phonons in carbon nanotubes

B.J. LeRoy, S.G. Lemay, J. Kong, and C. Dekker

*Kavli Institute of Nanoscience, Delft University of Technology, Lorentzweg 1, 2628 CJ, Delft, The Netherlands*

**The interplay between discrete vibrational and electronic degrees of freedom directly influences the chemical and physical properties of molecular systems. This coupling is typically studied through optical methods such as fluorescence, absorption, and Raman spectroscopy. Molecular electronic devices provide new opportunities for exploring vibration-electronic interactions at the single molecule level[1-6]. For example, electrons injected from a scanning tunneling microscope tip into a metal can excite vibrational excitations of a molecule in the gap between tip and metal[7]. Here we show how current directly injected into a freely suspended individual single-wall carbon nanotube can be used to excite, detect, and control a specific vibrational mode of the molecule. Electrons inelastically tunneling into the nanotube cause a non-equilibrium occupation of the radial breathing mode, leading to both stimulated emission and absorption of phonons by successive electron tunneling events. We exploit this effect to measure a phonon lifetime on the order of 10 nanoseconds, corresponding to a quality factor well over 10000 for this nanomechanical oscillator.**

Single-wall carbon nanotubes (SWCNTs) were grown by chemical vapor deposition on a Pt substrate, which had predefined 100 nm wide trenches etched in it. The SWCNTs were measured as-grown without any further processing. Details of the preparation of the samples used in this experiment have been reported previously[8]. Figure 1A sketches the setup for the measurements indicating the scanning tunneling microscope (STM) tip and the SWCNT crossing a trench. Figure 1B shows an STM topographic image of a nanotube suspended across a trench. The SWCNT is suspended over a distance of about 100 nm. Figure 1C is a zoom in on the tube in the region over the trench, demonstrating atomic resolution.

A STM can reveal information about electronic structure with high spatial resolution through tunneling spectroscopy. Figure 2A plots the normalized tunneling differential conductance, $(dI/dV)/(I/V)$, as a function of sample voltage near the edge of a trench for the semiconducting SWCNT of Fig. 1B. As the sample voltage is decreased, a series of sharp spikes are obtained due to the Coulomb staircase[8,9]. The spacing between peaks is a measure of the energy necessary to add an electron to the SWCNT. Figure 2B plots the normalized differential conductance at the center of the trench. Remarkably, new side peaks appear in addition to the previously observed Coulomb peaks. Such additional peaks in the differential conductance can occur when there are new channels through which electrons can tunnel onto the SWCNT. Figure 2C shows the full spatial dependence of the tunneling differential conductance. The four main Coulomb staircase peaks shift in energy due to a changing tip-SWCNT capacitance[8] but persist throughout the image. In contrast, the additional side peaks are localized in the part of the SWCNT that is suspended. That side peaks only occur in the suspended portion explains why they have not been observed in previous STM measurements where nanotubes are lying on a conducting surface.

An STM allows the resistance of one of the tunnel barriers to be changed, thus controlling the current through the SWCNT at a fixed applied voltage. Based on previous transport measurements, the resistance of the SWCNT-substrate barrier is known to be in the range 10-1000 kΩ. Since the resistance of the tip-SWCNT barrier (~1 GΩ) is much greater than the SWCNT-substrate barrier, the total resistance and tunnel rate is set by the tip-SWCNT barrier. Figure 3A plots the normalized differential conductance through a metallic SWCNT as a function of bias voltage for a low setpoint current, $I_{set}$ = 100 pA, and a sample voltage of -0.6 V. The peaks due to the Coulomb staircase appear in groups of four indicating that there are two spin-degenerate bands in this metallic SWCNT[10-12]. Figures 3B and C plot the differential conductance with a medium setpoint current of 300pA and a high current of 1000 pA, respectively. Remarkably, many additional peaks have appeared on either side of the main Coulomb staircase peaks. The peaks are equally spaced in energy on both sides of the main Coulomb peaks. While the side-peak intensities change, the energy relative to the Coulomb peak is found to be independent of the current.

We attribute these side peaks to phonon-assisted tunneling into the SWCNT. By absorbing or emitting a phonon, electrons tunneling onto the SWCNT can increase or decrease their energy by $\hbar\omega$, where $\omega$ is the frequency of the phonon. Side peaks thus appear at energies $\hbar\omega$ from the main Coulomb peaks. Figure 3D is a zoom in on one of the Coulomb peaks showing side peaks corresponding to both emission and absorption of a phonon, in direct analogy to the well-known Stokes and anti-Stokes peaks in Raman spectroscopy. The main Coulomb peak occurs when electrons tunnel from the STM tip to the SWCNT elastically (Fig. 3F). Peaks at energies above the Coulomb peaks (farther from $V$ = 0) are due to electrons emitting a phonon when tunneling (Fig 3E). Likewise, side peaks on the low-energy side of the Coulomb peaks (closer to $V$ = 0) correspond to electrons absorbing a phonon from the SWCNT when tunneling (Fig. 3G). Note that the observation of a strong peak for phonon absorption is very surprising given the fact that at equilibrium, without current, the population of phonons is extremely small at the low temperatures (5 K) of our experiment. The appearance of such peaks implies that current through the SWCNT induces a non-equilibrium phonon distribution. The phonon absorption effect has never been observed in electrical transport measurements. The non-equilibrium phonon distribution is furthermore responsible for the enhancement of the peaks on the high-energy side through stimulated emission.

It is possible to identify the specific phonon mode that we study in this experiment. For the 2.5 nm diameter SWCNT of Fig. 2, the energy is found to be E=11.8+/-1.4 meV, which corresponds to the energy of the radial breathing mode (RBM). To verify the origin of the side peaks as the radial breathing mode (RBM) phonon, we have repeated these measurements for a series of metallic and semiconducting SWCNTs. The measured energy of the side peaks as a function of inverse SWCNT diameter is plotted in Fig. 3H. The diameter of the SWCNTs is determined from the spacing of the first van Hove singularities[13]. The dashed line is the theoretical energy dependence of the RBM, $27.8/d$ meV, where $d$ is the diameter of the SWCNT in nm[14]. The measured data points are best fit with the equation 28.1+/-0.8 meV/$d$, which agrees very well with theory as well as with previously obtained results using Raman spectroscopy[14]. This demonstrates that the dominant phonon mode excited by electrons tunneling into the SWCNT is the RBM, in agreement with other recent observations[15]. Electrons that tunnel into the SWCNT move at the Fermi velocity, thus traveling 10 nm in ~10 fs while the period of the RBM oscillation is ~370 fs. This suggests that circumferentially symmetric and low-k values of phonons are preferentially excited. This explains why the RBM is observed while the many other modes in the phonon density of states are not. When considering the excitation of low-k phonons and bending modes, we can exclude acoustic phonons as these will not be observed because of their low energy.

The new peaks that are observed for increased values of the current allow us to estimate the lifetime of the phonon in an individual SWCNT. The fact that side peaks on the low-energy side are visible only for faster rates of electron tunneling implies that the decay time for the phonon mode in this SWCNT is longer than the average time between electron tunneling events at these rates. For example, for the case of Fig. 3B, the current for the first absorption peak visible at -0.15 V gives an average rate of electron tunneling events of 1.44 x $10^8$ $s^{-1}$. This gives a decay time of at least $\tau \cong 7$ ns, which is more than 50 times longer than observed in Raman experiments on bundles of SWCNTs[16]. The decay time in the Raman data may be caused by coupling to other SWCNTs or the substrate which are absent in our case. Our measured decay time corresponds to a lower bound for the quality factor $Q = \tau E/h$ of 20000 for the RBM. Typical values of Q that we obtain for the RBM in our devices are in the range 5000-30000. These large Q values are an order of magnitude larger than previous reports for other phonon modes in SWCNTs[17], and are on the same order as those found in lithographically fabricated nanomechanical oscillators[18].

To further quantify the effect of the tunneling current on the vibrational excitations of the SWCNT, we have compared the measured differential conductance as a function of current with a simple model of phonon-assisted tunneling. Figure 4A shows a specific differential-conductance peak as a function of voltage for a series of different currents. As the current is increased, the strength of the main Coulomb peak decreases whereas the peak associated with phonon emission increases. This shows that the probability of emitting a phonon is controlled by the rate of electrons passing through the SWCNT. The amplitude of these two peaks as a function of current is plotted in Fig. 4B showing the side peak becoming stronger than the main peak. Figure 4C and D plot the differential conductance as a function of voltage and current for a different SWCNT. As the current is increased, an increasing number of side peaks (up to 4) appears. This shows that multi-phonon excitations become possible at high current levels.

In analogy to photon-assisted tunneling, we can model the effect of the phonons as an oscillating potential on the SWCNT proportional to $\alpha \cos(\omega t)$, where $\alpha$ is the strength of the perturbation and $\omega$ is the frequency of the phonon mode[19]. This model predicts that the Coulomb peak at energy $\varepsilon_0$ evolves into a series of peaks $\sum_n J_n^2(\alpha) \frac{\partial f}{\partial V}(eV - \varepsilon_0 - n\hbar\omega)$, where $f$ is the Fermi function. $J_n$ is the nth order Bessel function of the first kind, and $n$ labels the peak number. The electron-phonon coupling term in the Hamiltonian is linear in the phonon annihilation and creation operators and therefore the strength of the perturbation $\alpha \propto \sqrt{N}$ where $N$ is the number of phonons. Because the tunnel current $I$ excites the phonons, a simple estimate is that $N \propto I$ and therefore $\alpha = \sqrt{I/\gamma}$, where $\gamma$ is an unknown fitting parameter. The dashed lines in Fig. 4B show the expected peak height for the main peak and the first excited state as a function of current using this model. This simple model captures the essential trends in the experimental results. A better quantitative agreement to the experimental results can be obtained however, if we assume that $N \propto I^2$. This gives a perturbation that is linear in the current, $\alpha = I/\beta$ where $\beta$ is an unknown fit parameter. The results of the simulations using this relationship for $\alpha$ are shown by the solid lines in Figure 4B and in Figure 4E. An excellent agreement is obtained with only 2 largely uncorrelated fitting parameters, the peak height at zero current and $\beta$. As the current is increased in both the experimental data and the simulations, the number and strength of side peaks associated with phonon-assisted tunneling increase, indicating that it is possible to control the population of phonons excited in a SWCNT.

Summing up, we have demonstrated that electrons passing through an individual carbon nanotube can populate phonon modes of the nanotube. This distribution can be probed with tunneling spectroscopy that shows that electrons can both absorb and emit phonons when tunneling onto the SWCNT. More theoretical work is needed to understand the current dependence of the phonon-assisted tunneling peaks. These new phenomena allow the measurement of long phonon lifetimes and high quality factors. Furthermore, it opens the door to the design of experiments with a known and controlled phonon population.

Acknowledgements

The authors are grateful to Ya. M. Blanter for helpful discussions. The authors would like to thank NWO and FOM for funding

Authors declare they have no competing financial interests

Correspondence and requests for materials should be addressed to C.D. (dekker@mb.tn.tudelft.nl)

Figure Captions

Figure 1 Measurement setup and topographic images. **A** Schematic diagram showing the setup for performing spectroscopy on suspended SWCNTs. A voltage is applied to the substrate with respect to the tip and the current flowing from the substrate through the SWCNT to the tip is measured. **B** STM image of a nanotube crossing a trench. The scale bar is 25 nm. The apparent width of the 2 nm-diameter tube is enlarged by tip convolution. **C** High-resolution image of the suspended portion of the SWCNT showing atomic resolution. The scale bar is 2 nm. The STM images were taken with a feedback current of 300 pA at -1 V. All of the measurements are performed at 5 K in an ultra-high-vacuum STM.

Figure 2 Spatially resolved spectroscopy along the suspended semiconducting SWCNT shown in Fig. 1B. **A** Spectroscopy along the blue line in **C**, showing four spikes associated with the Coulomb staircase. The Coulomb staircase behavior is determined by the two capacitances and resistances; the tip-SWCNT and the SWCNT-substrate. The peaks occur when the Fermi level of the substrate aligns with states in the conduction band of the SWCNT. **B** Spectroscopy along the red line in **C**, showing side peaks in addition to the Coulomb staircase peaks. **C** Plot of the normalized differential conductance (color scale) as a function of energy and position. Sharp spikes are visible at all positions due to the Coulomb staircase, while extra peaks are visible only in the center of the suspended region. The differential conductance was measured using lock-in detection with a 2 mV rms excitation voltage. The setpoint current was 300 pA at -1.25 V. The colored lines above **C** indicate the regions where the SWCNT is supported (black) and suspended (green).

Figure 3 Current and diameter dependence of phonon assisted tunneling. **A** Normalized differential conductance of a metallic SWCNT as a function of sample voltage taken with a low setpoint current $I_{set}$ at -0.6 V. The tip is located at the center of the suspended SWCNT. A series of sharp peaks is visible due to the Coulomb staircase as the Fermi level of the substrate aligns with unoccupied states of the SWCNT. **B**,**C** Same as **A** with increasing setpoint current. A series of side peaks has appeared near the main Coulomb staircase peaks due to absorption and emission of phonons. To convert from sample voltage to energy, the capacitances between the tip and SWCNT and the SWCNT and substrate must be known. The voltage dropped across the substrate-SWCNT junction is the applied voltage times the ratio of the tip-SWCNT capacitance and the total capacitance. These capacitances were determined from the spacing of the Coulomb peaks and the conductance between peaks[8,20]. **D** Zoom in on one of the peaks showing side peaks corresponding to the emission (**E**) and absorption (**G**) of phonons. The data shows a negative differential resistance between the main peak and the emission peak, which is often observed in these experiments. **E** Energy diagram for emission of a single phonon showing that an increased energy is needed for electron tunneling. The distance between the solid black line and the top of the black box represents the Coulomb charging energy. **F** Energy diagram for elastic tunneling where a level in the SWCNT is aligned with the leads. **G** Energy diagram for absorption of a phonon, which decreases the energy needed for tunneling. **H** Plot of the energy of the side peaks as a function of inverse diameter, showing a linear relationship. The horizontal error bars arise from the width of the van Hove singularities while the vertical ones are the standard deviation of the side-peak energy. The dashed line is the expected energy for the radial breathing mode obtained from Dresselhaus et al[14] and is plotted without any adjustable parameters.

Figure 4 Comparison of the observed current dependence with theory. **A** Normalized differential conductance versus energy for a series of different currents ranging from 5 to 50 pA. As the current is increased, the main peak at -0.10 V decreases while the side peak at -0.125 V, associated with emission of a phonon, increases. **B** Strength of the main (black squares) and first side peak (red triangles) as a function of the current at the given peak along with a fit using Bessel functions as described in the text. Dashed lines are for the perturbation $\alpha = \sqrt{I/\gamma}$ with $\gamma = 20$ pA while solid lines are for $\alpha = I/\beta$ with $\beta = 24.5$ pA. **C** Differential conductance as a function of bias voltage for a series of increasing currents from 500 to 1300 pA, for a different SWCNT than **A**. **D** Experimentally measured dependence of the differential conductance on current showing additional side peaks appearing as the current is increased. **F** Simulation of the effect of increasing the current using $\alpha = I/\beta$ with $\beta = 310$ pA, which changes the phonon population and allows multiple phonon excitations. The conductance for the nth peak is given by the square of the nth-order Bessel function of the first kind. The data for **D** and **E** concern the first peak of a semiconducting SWCNT with zero current in the gap, and therefore there are no side peaks associated with phonon absorption (i.e., on the right side of the main peak).

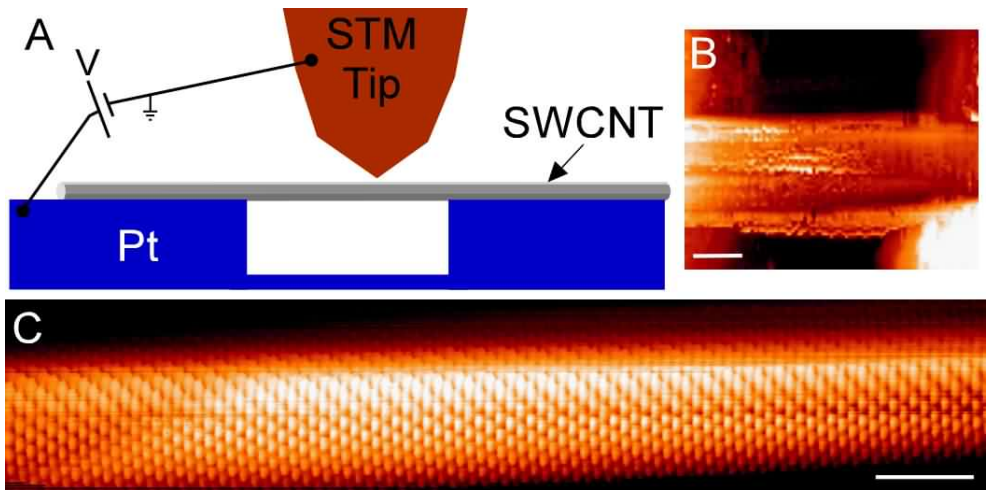

B.J. LeRoy et al, Figure 1

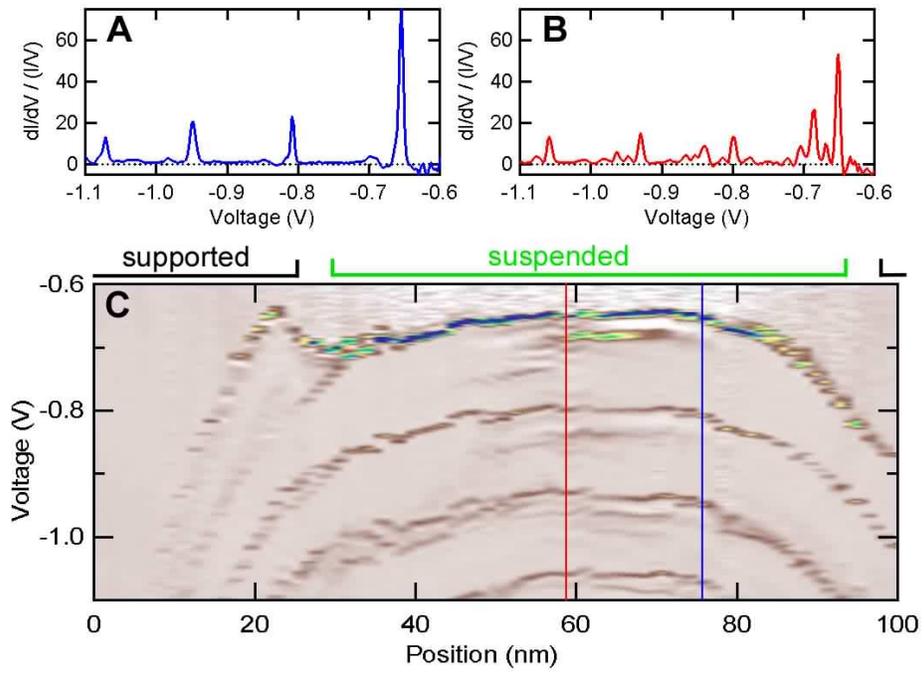

B.J. LeRoy et al, Figure 2

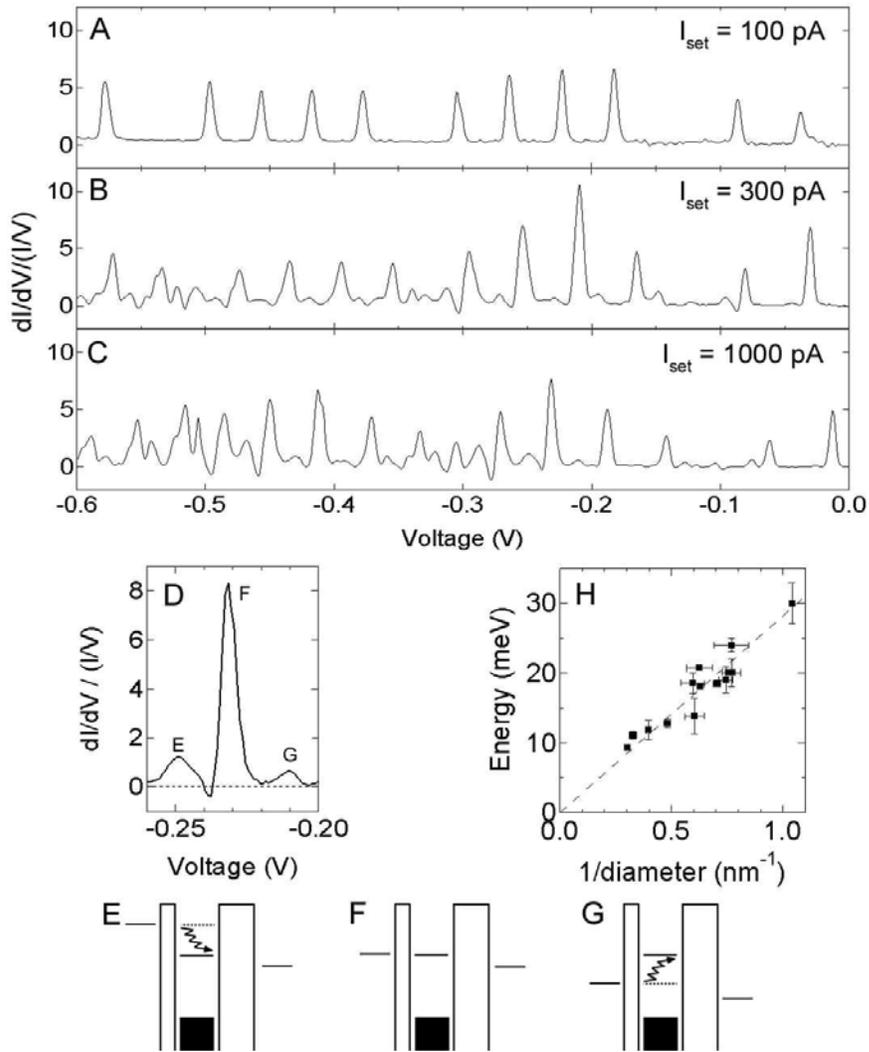

B.J. LeRoy et al, Figure 3

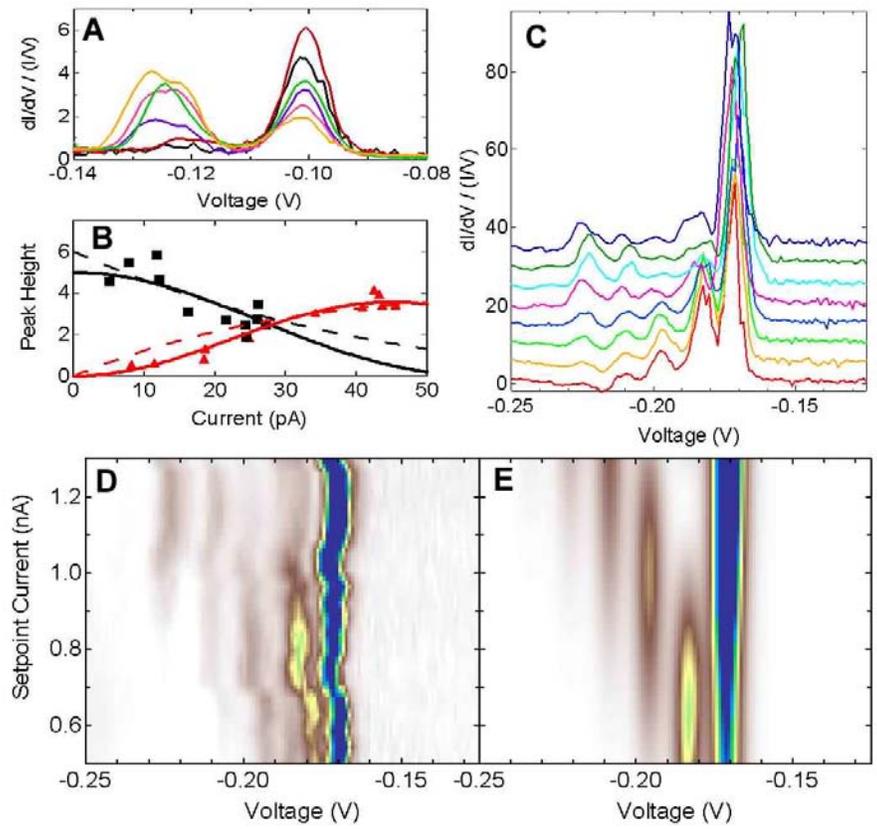

B.J. LeRoy et al, Figure 4